\documentclass[mathleft
]{an}
\usepackage{graphicx}
\usepackage{times}
\begin{document}

% The following seven commands are intended for editorial usage and should be ignored by
% the author(s).
\Pagespan{1}{}% Document's page range. 
% If second parameter is left empty, the last page is computed automatically.
\Yearpublication{}%
\Yearsubmission{}%
\Month{}%   
\Volume{}%  
\Issue{}% 
% \DOI{This.is/not.aDOI}% 

\title{Accretion disk winds in active galactic nuclei: X-ray observations, models and feedback}

\author{F. Tombesi\inst{1,2}\fnmsep\thanks{\email{francesco.tombesi@nasa.gov
      / ftombesi@astro.umd.edu}\newline}
%Example 
%for footnote, note the usage of the \texttt{fnmsep}
%command as separator between institute number and footnote mark} 
%\and  G.H. Ostwriter\inst{2,3}
}
\titlerunning{Accretion disk winds in active galactic nuclei}
\authorrunning{F. Tombesi}
\institute{
X-ray Astrophysics Laboratory and, NASA/Goddard Space Flight
Center, Greenbelt, MD 20771, USA
\and
Department of Astronomy, University of Maryland and CRESST, College Park, MD 20742, USA}

\received{}
\accepted{}
\publonline{later}


\abstract{Powerful winds driven by active galactic nuclei (AGN) are
  often invoked to play a fundamental role in the evolution of both
  supermassive black holes (SMBHs) and their host galaxies, quenching
  star formation and explaining the tight SMBH-galaxy relations. A
  strong support of this “quasar mode” feedback came from the recent
  X-ray observation of a mildly relativistic accretion disk wind in a
  ultraluminous infrared galaxy (ULIRG) and its connection with a
  large-scale molecular outflow, providing a direct link between the
  SMBH and the gas out of which stars form. Spectroscopic
  observations, especially in the X-ray band, show that such accretion
  disk winds may be common in local AGN and quasars. However, their
  origin and characteristics are still not fully understood. Detailed
  theoretical models and simulations focused on radiation,
  magnetohydrodynamic (MHD) or a combination of these two processes to
  investigate the possible acceleration mechanisms and the dynamics of
  these winds. Some of these models have been directly compared to
  X-ray spectra, providing important insights into the wind
  physics. However, fundamental improvements on these studies will
  come only from the unprecedented energy resolution and sensitivity
  of the upcoming X-ray observatories, namely ASTRO-H (launch date early 2016) and Athena (2028).}

\maketitle

\section{Introduction}

Most galaxies host a supermassive black hole (SMBH) at their center,
with masses ranging from a few million (e.g., for the one in our Milky
Way) up to several billion times the one of the Sun. Surprisingly,
the SMBH mass is found to correlate with several properties of the host
galaxy. For instance, galaxies hosting more massive black holes also
possess more massive bulges that contain on average faster-moving
stars (e.g., Magorrian et al. 1998; Ferrarese \& Merritt 2000). This suggests some sort of feedback mechanism(s) between a galaxy's black hole and the star-formation process. Yet there is still no adequate explanation for how a black hole's activity, which in principle may affect only a region of few times the size of our solar system, could actually influence a whole galaxy, which encompasses regions roughly a billion times larger. 

During their active phases, accreting SMBHs can inject significant
amounts of energy in their surroundings and they are called active
galactic nuclei (AGN). Therefore, observations of AGN provide the
key to directly study the feedback phenomenon in action. Radiation and
jets from AGN can indeed interact with the interstellar medium
leading to ejection or heating of the gas. However, they alone are not
able to establish the intricate accretion/ejection feedback cycle
linking the central SMBH to its host galaxy (e.g., Fabian 2012; Pounds
2014; King \& Pounds 2015; Combes 2015). Increasing evidence point toward another promising
player in AGN feedback: accretion disk winds. Observations in the
X-ray band are particularly important because this radiation is
produced very close to the central SMBH and it is energetic enough to
pass through dense layers of absorbing material, therefore retaining a
wealth of information about the AGN and the host galaxy environment. 

This brief review paper is organized as follows: in \S2 we describe the X-ray observations of disk
winds, in \S3 we compare the models that have been proposed for their
characterization, in \S4 we discuss the AGN feedback from these disk
winds, and in \S5 we show some of the fundamental improvements expected from upcoming X-ray observatories.

\section{X-ray observations of disk winds}

The first evidence of ionized absorption in the X-ray spectrum of an
AGN was reported by Halpern (1984) comparing two spectra of the quasar
MR~2251$-$178 taken in 1979 and 1980 with the \emph{Einstein}
satellite. The data showed a change in the soft X-ray band, which was
interpreted as an ionized cloud crossing the continuum source.  

The advent of the higher energy resolution grating spectrometers onboard \emph{Chandra}
and \emph{XMM-Newton} provided a revolution in these studies, showing that the
absorption is be composed of a number of lines and edges from different elements at different ionization
states (e.g., Kaspi et al.~2002). Importantly, the energies of these lines
are found to be systematically blue-shifted compared to the expected
values, indicating that the material is likely a wind outflowing from the central
regions of these galaxies with velocities in the range of
$v_{out}$$\sim$100--1,000 km~s$^{-1}$ (e.g., Kaastra et al.~2000; McKernan etal~2007;
Gofford et al.~2011; Lobban et al.~2011; Detmers et al.~2011; Reeves et al.~2013; Kaastra et
al.~2014). Overall, such warm absorbers (WAs)
are detected in more than half of local Seyfert galaxies and have
ionization and column densities in the range log$\xi$$\sim$1--3 erg~s$^{-1}$~cm and
$N_H$$\sim$$10^{20}$--$10^{22}$ cm$^{-2}$, respectively (e.g., Crenshaw \& Kraemer
2012). 

A more extreme type of outflows, often referred to as ultrafast
outflows (UFOs), are instead observed mostly in the Fe K band through
blue-shifted Fe~XXV and Fe~XXVI absorption lines (e.g. Chartas et
al.~2002; Pounds et
al.~2003; Dadina et al.~2005; Reeves et al.~2009; Cappi et al.~2009; Tombesi et al. 2010a,
b; Giustini et al.~2011; Gofford et al. 2013; Tombesi et al.~2014; Nardini et al.~2015; Tombesi et al.~2015; but see also Gupta et
al~2013, 2015 for UFO detections in the soft X-rays). The UFOs are highly
ionized, with ionization parameter log$\xi$$\sim$4--6
erg~s$^{-1}$~cm. They can have high column densities in the range
$N_H$$\sim$$10^{22}$--$10^{24}$ cm$^{-2}$. Most importantly, the
implied outflow velocities are often mildly relativistic, in the range
$v_{out}$$\sim$0.03--0.3~c, where c is the speed of light (e.g., Tombesi et
al.~2011; Gofford et al. 2013). 

The large dynamic range obtained when considering the WAs and UFOs
together, spanning several orders of magnitude in ionization, column density,
velocity and distance suggests that the closer is the absorber to the
central black hole, the higher are the values of these parameters and
consequently the mechanical power. This suggests that these absorbers could represent parts of a same
large-scale outflow observed at different locations from the black
hole. The UFOs are likely launched from the inner accretion disk and
the WAs at larger distances, of the order of pc-scales or larger
(e.g. Blustin et al.~2005; Tombesi et al.~2012a, 2013; Gofford et al.~2015).

\section{Models of disk winds}

The detailed acceleration mechanisms of disk winds in AGN are still
matter of intense research. However, they can be classified in
thermal, radiation and magnetohydrodynamic (MHD), depending on their
main driving force. 

Thermal-driven winds are accelerated by the thermal pressure of the
gas and they can reach a maximum velocity of
$\sim$1,000 km~s$^{-1}$. They are ejected from the outer accretion
disk or obscuring torus at $\sim$pc scales from the central SMBH. Some of the WA
components in Seyferts may be associated with such winds (e.g.,
Chelouche \& Netzer 2005). 

Given the intense radiation field in most AGN, radiation pressure itself can be a very effective way to
drive a disk wind (e.g., Elvis 2000; Sim et al.~2008, 2010, 2012;
Ohsuga et al.~2009). 
The main source of opacity is due to UV absorption
lines (e.g., Proga, Stone \& Kallman 2000; Higginbottom et al.~2014). Compton scattering can
also be effective if the luminosity of the source is close to
Eddington (e.g., King \& Pounds 2003). The wind is ejected from
different locations on the accretion disk and can be accelerated to
very high velocities, up to values of $v_{out}$$\sim$0.1c. This
provides a very promising explanation for the observations of UFOs in
luminous AGN. 

For instance, a case study was recently reported for PDS~456 by Hagino
et al.~(2015), see Fig.~1. Using the \emph{MONACO} radiation transfer code, they
have been able to model the blue-shifted Fe XXV--XXVI absorption lines
due to the disk wind in the \emph{Suzaku} spectra of this source. They
estimanted fundamental parameters of the wind, such as an
inner launching radius of $\sim$10$r_s$ ($r_s = 2GM_{BH}/c^2$), inclination
angle of $\sim$48$^{\circ}$ and an outflow velocity of $\sim$0.3c. The
higher the inclination angle, the higher is the EW of the line due to
the inclrease in column density. Instead, the inner radius mostly
affects the velocity centroid and shape of the line. Thus, the smaller is the radius, the higher is
the velocity and the broader is the high energy wing of the line
(e.g., Fukumura et al.~2015).

\begin{figure}
\includegraphics[width=80mm,height=65mm]{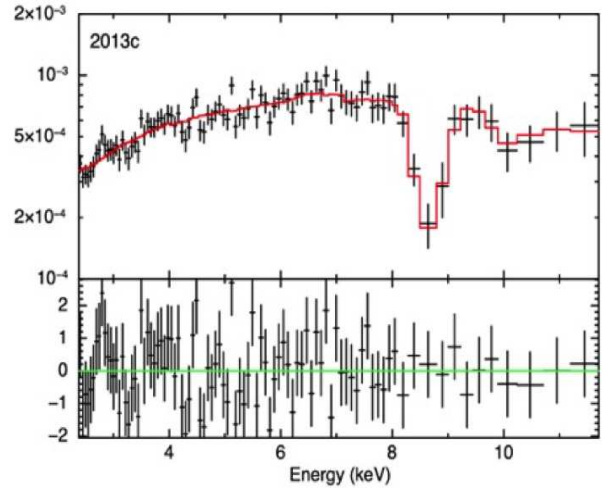}
\caption{Best-fit radiation-driven wind model of one \emph{Suzaku}
  spectrum of PDS~456 observed in 2013, from Hagino et al.~(2015).}
\label{label1}
\end{figure}

Another promising possibility to explain the origin of disk winds in
AGN (and possibly also X-ray binaries) is offered by MHD
models (e.g., Kazanas et al.~2012). Magnetic fields are fundamental for the onset of accretion and
the formation of relativistic jets, therefore it is plausible that they may have an effect
in the production of disk winds as well. In this case the wind is
accelerated by the centrifugal force of the magnetic field lines
anchored on the disk and the magnetic pressure (e.g., Blandford \& Payne
1982; Konigl \& Kartje 1994; Proga 2000; Everett \& Ballantyne 2004; Everett 2005; Fukumura et al.~2010a). The wind is
ejected from different regions of the disk. This causes a stratification, with increasing column density, ionization and velocity
closer to the central SMBH. In particular, the resultant wind velocity
is proportional to the disk rotational velocity at each radius and it
can reach up to relativistic values (e.g., Fukumura et al. 2010b,
2014). The wind possibly originates from just a few $r_s$ from the
SMBH up to the outer disk or torus.

\begin{figure}
\includegraphics[width=80mm,height=65mm]{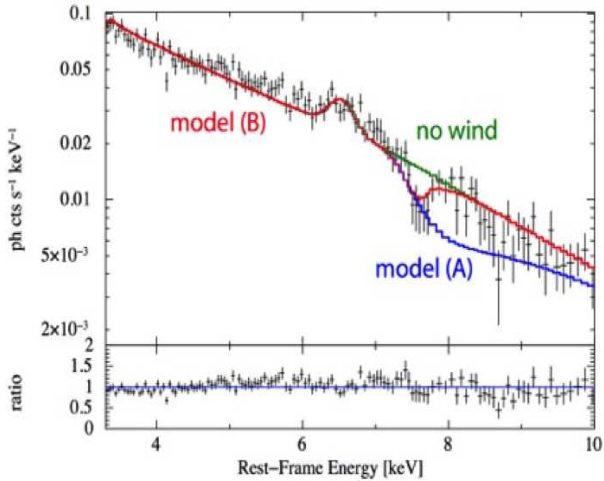}
\caption{Best-fit MHD-driven wind model of the 2001 \emph{XMM-Newton}
  spectrum of PG~1211$+$143, from Fukumura et al.~(2015). The green,
  blue, and red lines indicate the case without wind model, with inner
wind radius fixed to the innermost stable circular orbit, and the best-fit
model with an inner wind radius of $\simeq$30$r_s$, respectively. }
\label{label2}
\end{figure}

The first MHD modeling of an AGN disk wind observed in the X-rays was recently reported by
Fukumura et al.~(2015) for the quasar PG~1211$+$143, see Fig.~2. Using the
\emph{mhdwind} model in XSPEC the authors have been able to
successfully parameterize the blue-shifted Fe~K absorption line in the 2001
\emph{XMM-Newton} spectrum of this source. They estimate an inner wind
launching radius of $\sim$30$r_s$, an inclination angle of
$\sim$50$^{\circ}$ and an outflow velocity of
$\simeq$0.1--0.2c. 

These models provide a very good starting point for the study of disk
winds in AGN. However, it is important to note that the real situation can be much
more complex. In fact, accretion and ejection physics
should be considered together in a self-consistent way. Therefore, the
accretion disk, the wind and the jet should be studied simultaneously (e.g.,
Ohsuga et al.~2009; Tchekhovskoy et al.~2011; Sadowski et al.~2013,
2014; McKinney et al.~2014; Yuan \& Narayan 2014). Moreover, the different
acceleration mechanisms can also be present at the same time or some
of them may dominate for certain wind launching radii. Bright radio galaxies are promising candidates for
these studies, as they simultaneously show the disk, the wind and the jet
(e.g., Marscher et al.~2002; Tombesi et al.~2012b, 2014). 

\section{AGN wind feedback}

There are several indications of relations between SMBHs and their host galaxies. For instance,
the M$_{BH}$--$\sigma$ relation shows that the bigger the mass of the SMBH,
the higher is the velocity dispersion of stars in the galaxy bulge
(e.g., Kormendy \& Ho 2013). Moreover,
large-scale computer simulations of galaxy evolution show that the high mass end of the galaxy stellar mass
function is over-predicted and some phenomena should be responsible
for the quenching of the star formation. One promising possibly being feedback from the central SMBH (e.g., Bower et al.~2012). 

\begin{figure}
\includegraphics[width=80mm,height=65mm]{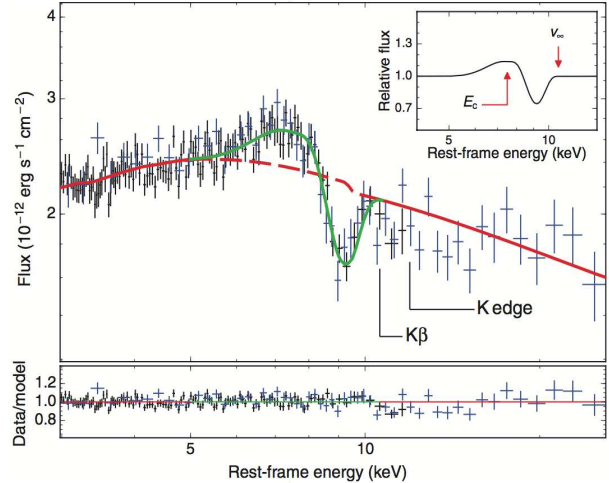}
\caption{Simultaneous \emph{XMM-Newton} and \emph{NuSTAR} spectra of PDS~456 in
  the Fe K band showing a clear P-Cygni profile from a disk wind with
  velocity $\simeq$0.3c, from Nardini et al.~(2015).}
\label{label3}
\end{figure}

Therefore, a fundamental question is, \emph{how do SMBHs affect galaxy evolution?}
The typical radius of a SMBH is $\sim$10$^{9}$ times smaller than
that of a galaxy. Considering the volume, the SMBH is $\sim$10$^{27}$
times smaller! Moreover, the typical mass of a SMBH is only $\sim$1\%
of the stellar bulge mass (e.g., Magorrian et al.~1998). However, it is
important to note that there is an huge amount of gravitational energy
``released'' by SMBHs, which in total can be comparable to the binding energy of the
entire galaxy bulge! Therefore, if there is a way to tap into this
energy, SMBHs may have a strong impact on the host galaxy. 

Indeed, the conversion of gravitational energy into radiation or jets
and winds can have a profound impact on the AGN host galaxy. Here, we
focus more on the winds, which are thought to be responsible for the
so-called ``quasar-mode feedback'' (e.g., Silk \& Rees 1998; Fabian 2012; King
Pounds 2015). In this regard,
two recent papers reported compelling evidence for the presence of powerful disk winds
in quasars. 

Combining simultaneous X-ray spectra collected with \emph{XMM-Newton}
and \emph{NuSTAR}, Nardini et al.~(2015) showed the clear presence of
a P-Cygni line profile in the Fe K band of the luminous quasar
PDS~456, see Fig.~3. This is due to a powerful, large opening angle disk wind with
an outflow velocity of $v_{out}$$\sim$0.3c. This X-ray wind is observed
close to the accretion disk, but it is very likely to have a
strong effect at much larger scales in the host galaxy.

On the other hand, Tombesi et al.~(2015) reported, for the first time,
the connection between a powerful accretion disk wind detected in the
X-ray band with \emph{Suzaku} and a large-scale molecular outflow
detected in the IR with \emph{Herschel} in IRAS~F11119$+$3257, see Fig.~4. 
This is a local ($z = 0.189$) ULIRG hosting a quasar in the center with a luminosity of
$L$$\simeq$$10^{46}$~erg~s$^{-1}$ (Veilleux et al.~2013). This source, as most ULIRGs,
is likely the result of a previous merger between two
galaxies. Therefore, the study of this type of objects is very important for the connection with
galaxy evolution and whether the merger process is an effective way to
feed the central SMBH to very high rates (within the uncertainties, this source is accreting at $\sim$5 times the
Eddington limit).

\begin{figure}
\includegraphics[width=80mm,height=65mm]{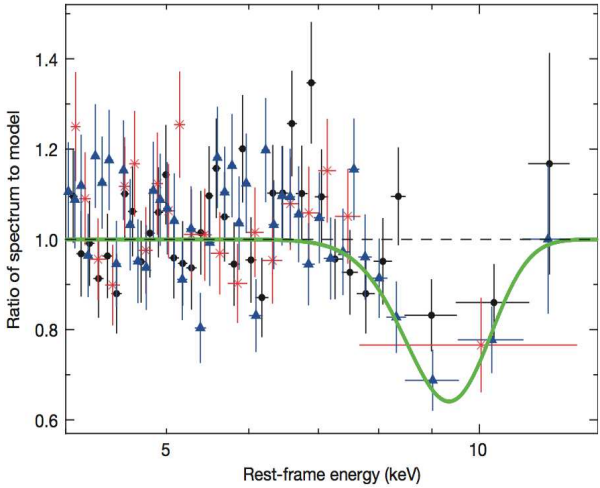}
\caption{Ratio between the \emph{Suzaku} XIS0 (black), XIS1 (red) and
  XIS3 (blue) spectra of IRAS~F11119$+$3257 in the Fe K band showing a
  clear broad and blue-shifted absorption line from a disk wind with
  velocity of $\simeq$0.25c, from Tombesi et al.~(2015).}
\label{label4}
\end{figure}

 The OH 119$\mu$m line profile in IRAS~F11119$+$3257 shows
a prominent P-Cygni profile, indicating a molecular outflow with maximum
velocity of 1,000~km~s$^{-1}$ at a scale of $>$300~pc from the SMBH. 
This corresponds to a very high mass outflow rate of
$\sim$800~M$_{\odot}$~yr$^{-1}$.  
The X-ray spectrum shows a broad and blue-shifted absorption line at
the energy of $\simeq$9~keV. The best-fit XSTAR model (Kallman \&
Bautista 2001) indicates 
a highly ionized wind with outflow
velocity $v_{out}$$\simeq$0.255c, column density $N_H$$\simeq$$6\times
10^{24}$~cm$^{-2}$ and covering fraction $>$0.85. 
The mechanical energy of the disk wind and the molecular outflow are
log$\dot{E}_K$$=45.4^{+0.4}_{-0.5}$ erg~s$^{-1}$ and
log$\dot{E}_K$$=44.4\pm0.5$ erg~s$^{-1}$, respectively. These
corresponds to $\simeq$15\% and $\simeq$3\% of the AGN luminosity, respectively. The
two values are consistent considering an efficiency $f$$=$$0.22\pm0.07$ derived
from the ratios of the two convering fractions. This result is in
agreement with the "quasar-mode feedback" model in which the mildly-relativistic disk
wind produces a strong shock in the interstellar
medium, which then expands adiabatically as a hot bubble and drives
the molecular outflow at galaxy-scales (e.g., Zubovas \& King 2012; Faucher-Gigu\`ere \&
Quataert 2012; Wagner et al.~2013; Zubovas \& Nayakshin 2014).

\section{Future missions}

Currently, only the CCD instruments onboard \emph{XMM-Newton}, \emph{Suzaku} and
\emph{NuSTAR} have provided enough sensitivity to study the crucial Fe K band in detail. In particular,
they allowed the detection of UFOs at E$>$7~keV in local AGN and
to estimante their parameters. However, there are still
uncertainties, mostly from the fact that the width of the lines is
often unconstrained and the limited sensitivity to higher order Fe K line
features. The improved energy resolution of the microcalorimeters onboard
\emph{ASTRO-H} and \emph{Athena} will allow to study such Fe K
features with unprecedented detail. 

\begin{figure}
\includegraphics[width=80mm,height=65mm]{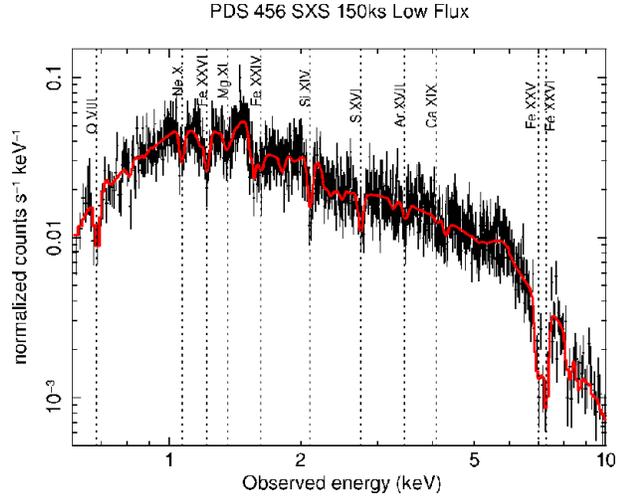}
\caption{Simulated \emph{ASTRO-H} microcalorimeter spectrum of PDS~456 in the interval
     E$=$0.5-10~keV. A 150ks exposure and the low flux case are
     assumed. The vertical dotted lines indicate the expected
     blue-shifted absorption lines from the UFO with velocity
     $v_{out}$$\simeq$0.23c. Both Fe K and several lower-Z element
     lines will likely be detected with high significance.}
\label{label5}
\end{figure}

The \emph{ASTRO-H} X-ray observatory (Takahashi et al.~2014), with a launch date planned for
early 2016, will have a microcalorimenter with a unprecedented
combination of high energy resolution of $\simeq$5~eV and high
sensitivity in the broad energy range between E$=$0.5--10~keV. Moreover, \emph{ASTRO-H} will have three
other instruments, which will allow to simultaneously cover whole
X-ray band, from 0.5~keV up to 500~keV.

As a case study we consider the simulated \emph{ASTRO-H}
observation of PDS~456 (for more details, see the AGN winds White
Paper by Kaastra et al.~2014). This is one of the quasars with the most powerful UFO
detected in the X-rays. The UFO in PDS~456 
was observed having a variable column density in the range
$N_H$$\simeq$$1\times 10^{23}$--$8\times 10^{23}$~cm$^{-2}$. The
column density appears to be higher for lower flux states. We assume
two representative cases, a high flux ( $3.6\times
10^{-12}$~erg~s$^{-1}$~cm$^{-2}$ in the 2--10~keV) / low column
($N_H=1\times 10^{23}$~cm$^{-2}$) and a low flux ($1.5\times
10^{-12}$~erg~s$^{-1}$~cm$^{-2}$ in the 2--10~keV) / high column
($N_H=5\times 10^{23}$~cm$^{-2}$), respectively. We assume the baseline
model of Gofford et al.~(2014). The exposure time
required to detect the UFO at $\ge$6$\sigma$ significance in both
low and high flux cases is 150~ks.

\begin{figure}
\includegraphics[width=80mm,height=65mm]{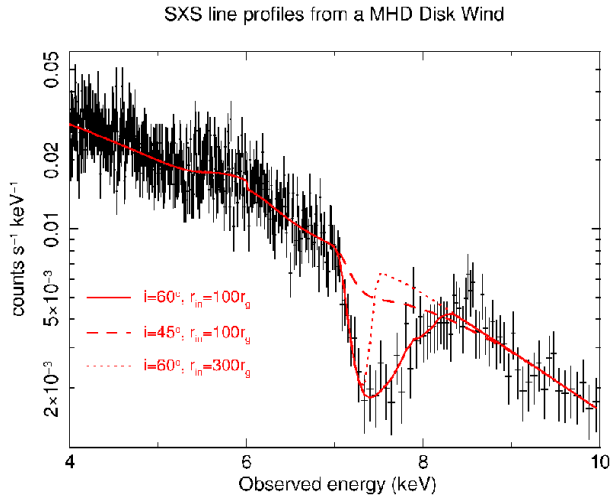}
\caption{Simulated \emph{ASTRO-H} microcalorimeter spectrum of PDS~456 in the Fe K band
     (E$=$4--10~keV). The lines illustrate the simulated MHD
     disk wind profiles using the model of Fukumura et al.~(2010) and
     the high flux continuum of PDS~456. Three representative profiles are shown for
     different values of the wind inclination angle ($i$) and
     launching radius ($r_{in}$). The microcalorimeter will
     allow to explore the geometry and acceleration of the wind.}
\label{label6}
\end{figure}

The microcalorimeter observation will allow to estimate the UFO parameters with
unprecedented accuracy ($N_H$ at 5-15\%,
log$\xi$ at 1-5\% and $v_{out}$ at 0.5-1\%). This is fundamental to decrease the uncertainties on the mass outflow rate and mechanical power of these winds and therefore to quantify the impact of the SMBH on the host galaxy through AGN feedback. From Fig.~5 we see that several other absorption lines due to lower-Z
elements at E$<$6~keV (mainly H/He-like Ar, Ca, S, Si, Ne, O, Mg) may
likely be observed in the spectrum. This will provide an additional strong
support for the existence of the UFO than the Fe XXV-XXVI lines alone. 

Moreover, for the first time, the microcalorimeter will allow to
resolve the line width with an accuracy of less than 10\%. This will allow to measure
the turbulent velocity of the plasma. Moreover, the two Fe~XXV He$\alpha$ and Fe~XXVI
Ly$\alpha$ lines will be distinguishable. Thus, the \emph{ASTRO-H}
observation will allow to explore 
different line broadening mechanisms. It will also 
be possible to investigate a distinction between radiation pressure
and MHD acceleration mechanisms (e.g., Sim et al.~2008; Fukumura et
al.~2010). In fact, the shape of the magnetic field lines in MHD winds
and the more equatorial
geometry of the radiation driven case will strongly affect the
absorption line profile (e.g., Giustini \& Proga 2012). In Fig.~6 we show the
representative case of an MHD disk wind profile using the model of
Fukumura et al.~(2010). Three representative profiles are shown for
     different values of the wind inclination angle ($i$$=$45$^{\circ}$--60$^{\circ}$) and
     launching radius ($r_{in}$$=$100--300~$r_g$). 

Another case study is the \emph{ASTRO-H} observation of the radio
galaxy 3C~120 (for more details, see the Broad-band White Paper by
Coppi et al.~2014). This is one of the most relevant and promising
targets for exploring disk winds in the context of the disk-jet
coupling. Previous X-ray observations, augmented by the
high-resolution radio data, hinted at a close link between the inner disk
state transitions, the jet formation processes and possibly also UFO
ejections (e.g., Marscher et al.~2002; Chatterjee et al.~2009; Tombesi
et al.~2012b; Lohfink et al.~2013).

\begin{figure}
\includegraphics[width=80mm,height=65mm]{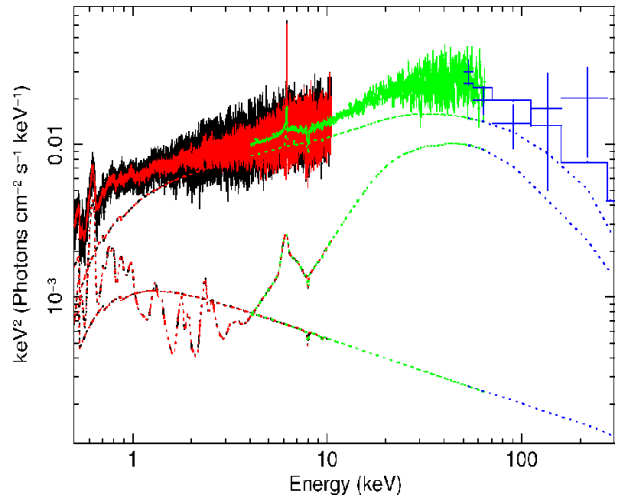}
\caption{Simulated 100\,ks \emph{ASTRO-H} broad-band spectrum of
  3C\,120 in the energy range $0.5-300$\,keV, corresponding to the
  state in which the jet knot is ejected and the inner disk is
  disrupted (the inner disk radius recedes from $r_{in} \simeq r_g$ up
  to $r_{in} \simeq 38\,r_g$). The microcalorimeter SXS spectrum is shown in black, the SXI in red, the HXI in green, and the SGD in blue.}
\label{label7}
\end{figure}

We consider the combined disk-jet model of Lohfink et al.~(2013), in which the jet X-ray emission is parameterized with a steep
power-law with the photon index of $2.5-4$, while the disk (disk
corona) continuum is represented by a power-law component with the
photon index of $1.7-2.4$ and the high energy cut-off at $150$\,keV.
In the first case, the disk extends down to the ISCO. This
is modeled in XSPEC with a relativistically blurred ionized reflection
component with $q\simeq 7$, inclination $i \simeq 15^{\circ}$, and the
ionization $\xi \simeq 200$\,erg\,s$^{-1}$\,cm. The neutral distant
reflector (pexmon) is included with a reflection fraction $R \simeq
2$. In the second case, which is related to the launch of the jet
knot, the inner disk is disrupted and the inner radius recedes to
$r_{\rm in} \simeq 38\,r_g$. The relativistic blurred reflection has
now a more standard value of $q \simeq 3.5$ and the neutral
reflection fraction is $R \simeq 0.26$. In this latter case, a UFO with
the velocity $v_{out} \simeq 0.16\,c$ was detected by Tombesi et
al.~(2014). This outflow is modeled with an XSTAR table assuming a
turbulent velocity of 3,000\,km\,s$^{-1}$, ionization $\log \xi \simeq
4.9$\,erg\,s$^{-1}$\,cm, and column density $N_{\rm H} \simeq 5 \times
10^{22}$\,cm$^{-2}$. If modeled with an inverted Gaussian absorption
line, this is equivalent to the energy of $E \simeq 8.23$\,keV, line
width of $\sigma \simeq 110$\,eV, and EW\,$\simeq -20$\,eV. 

The simulated 100~ks \emph{ASTRO-H} broad-band spectrum of 3C\,120 
is shown in Fig.~7. The source will be detected at high
significance with all the four instruments, from 0.5\,keV up to
300\,keV. This will allow to simultaneously constrain the corona
continuum, the jet emission, and the neutral/ionized reflection. In
addition, the inner disk radius will be measured with 15\%
accuracy. Moreover, the presence of the disk wind will be determined
by means of the detection of the Fe K UFO. In combination with an
multiwavelength monitoring, will allow for a direct comparison between the activity level of the intermittent jet, the state of the accretion disk, and the energetics of the disk wind. 

The very high effective area (more than $\sim$10x of \emph{ASTRO-H}) of the microcalorimeter onboard the
future (launch date 2028) \emph{Athena} X-ray observatory
(Nandra et al.~2013) will allow to extend the study of AGN wind feedback to higher redshifts
($z\simeq$2--3), where the peak of the AGN population is
expected. Moreover, it will be possible to study not only the
kinematics but also the dynamics of disk winds in local Seyfert
galaxies, thereby directly comparing the data with detailed simulations
(Cappi et al.~2013). 

\section{Conclusions}

Powerful winds driven by active galactic nuclei can play a fundamental role in the evolution of both
  supermassive black holes and their host galaxies, quenching
  star formation and explaining the tight SMBH-galaxy relations. A
  strong support of this “quasar mode” feedback comes from X-ray
  observation of mildly relativistic accretion disk winds in active
  galaxies.The unprecedented energy resolution and sensitivity
  of the upcoming X-ray observatories, namely \emph{ASTRO-H} and \emph{Athena}, are
  expected to provide revolutionary improvements in this field.

%\begin{figure}
%\includegraphics[width=80mm,height=65mm]{plot_3C120_SXS.ps}
%\caption{\emph{ASTRO-H} microcalorimeter SXS spectrum of 3C~120 in the Fe K
%  band, between E$=$5.5--8.5~keV. Emission and absorption lines will
%  be detected with very high accuracy.}
%\label{label1}
%\end{figure}

\acknowledgements
FT would like to thank Chris Done and Norbert Schartel for
organizing the very interesting and productive workshop ``The Extremes
of Black Hole Accretion'' held at ESAC, Madrid, Spain on June 8-10
2015.

%\newpage%%%%%%%%%%%%%%%%%%%%%%%%%%%%%%%%%%%%%%%%%%%%%%%%%%%%%%


\begin{thebibliography}{}
\bibitem[Blandford \& Payne(1982)]{1982MNRAS.199..883B} Blandford,
  R.~D., \& Payne, D.~G.\ 1982, \mnras, 199, 883 
\bibitem[Bower et al.(2012)]{2012MNRAS.422.2816B} Bower et al. \ 2012,
  \mnras, 422, 2816 
\bibitem[Blustin et al.(2005)]{2005A&A...431..111B} Blustin et al.\ 2005, A\&A, 431, 111 
\bibitem[Cappi et al.(2009)]{2009A&A...504..401C} Cappi et al.\ 2009,
  A\&A, 504, 401
\bibitem[Cappi et al.(2013)]{2013arXiv1306.2330C} Cappi, M., Done, C., Behar, E., et al.\ 2013, arXiv:1306.2330 
\bibitem[Chartas et al.(2002)]{2002ApJ...579..169C} Chartas, G. et
  al.\ 2002, \apj, 579, 169 
\bibitem[Chatterjee et al.(2009)]{2009ApJ...704.1689C} Chatterjee, R.,
  et al.\ 2009, ApJ, 704, 1689 
\bibitem[Chelouche \& Netzer(2005)]{2005ApJ...625...95C} Chelouche,
  D., \& Netzer, H.\ 2005, \apj, 625, 95 
\bibitem[Combes(2015)]{2015IAUS..309..182C} Combes, F.\ 2015, IAU
  Symposium, 309, 182 
\bibitem[Coppi et al.(2014)]{2014arXiv1412.1190C} Coppi et al.\ 2014,
  arXiv:1412.1190
\bibitem[Crenshaw \& Kraemer(2012)]{2012ApJ...753...75C} Crenshaw,
  D.~M., \& Kraemer, S.~B.\ 2012, \apj, 753, 75 
\bibitem[Dadina et al.(2005)]{2005A&A...442..461D} Dadina et al. \ 2005, A\&A, 442, 461
\bibitem[Detmers et al.(2011)]{2011A&A...534A..38D} Detmers et al.\
  2011, A\&A, 534, A38
\bibitem[Elvis(2000)]{2000ApJ...545...63E} Elvis, M.\ 2000, \apj, 545,
  63  
\bibitem[Everett 
\& Ballantyne(2004)]{2004ApJ...615L..13E} Everett, J.~E., \&
Ballantyne, D.~R.\ 2004, \apjl, 615, L13 
\bibitem[Everett(2005)]{2005ApJ...631..689E} Everett, J.~E.\ 2005,
  \apj, 631, 689 
\bibitem[Fabian(2012)]{2012ARA&A..50..455F} Fabian, A.~C.\ 2012,
  \araa, 50, 455 
\bibitem[Faucher-Gigu{\`e}re \& Quataert(2012)]{2012MNRAS.425..605F}
  Faucher-Gigu{\`e}re, C.-A., \& Quataert, E.\ 2012, \mnras, 425, 605 
\bibitem[Ferrarese \& Merritt(2000)]{2000ApJ...539L...9F} Ferrarese,
  L., \& Merritt, D.\ 2000, \apjl, 539, L9
\bibitem[Fukumura et al.(2010)]{2010ApJ...715..636F} Fukumura, K., et
  al.\ 2010a, \apj, 715, 636 
\bibitem[Fukumura et al.(2010)]{2010ApJ...723L.228F} Fukumura, K., et al.\ 2010b, \apjl, 723, L228 
\bibitem[Fukumura et al.(2014)]{2014ApJ...780..120F} Fukumura et al.\
  2014, \apj, 780, 120 
\bibitem[Fukumura et al.(2015)]{2015ApJ...805...17F} Fukumura et al.\
  2015, \apj, 805, 17 
\bibitem[Giustini et al.(2011)]{2011A&A...536A..49G} Giustini et al.\
  2011, A\&A, 536, A49 
\bibitem[Giustini \& Proga(2012)]{2012ApJ...758...70G} Giustini, M., \& Proga, D.\ 2012, ApJ, 758, 70 
\bibitem[Gofford et al.(2011)]{2011MNRAS.414.3307G} Gofford et al.\
  2011, \mnras, 414, 3307
\bibitem[Gofford et al.(2013)]{2013MNRAS.430...60G} Gofford et al. \
  2013, \mnras, 430, 60
\bibitem[Gofford et al.(2014)]{2014ApJ...784...77G} Gofford et al.\ 2014, \apj, 784, 77 
\bibitem[Gofford et al.(2015)]{2015MNRAS.451.4169G} Gofford et al.\ 2015, \mnras, 451, 4169 
\bibitem[Gupta et al.(2013)]{2013ApJ...772...66G} Gupta et al.\ 2013, \apj, 772, 66 
\bibitem[Gupta et al.(2015)]{2015ApJ...798....4G} Gupta et al.\ 2015,
  \apj, 798, 4
\bibitem[Hagino et al.(2015)]{2015MNRAS.446..663H} Hagino et al.\
  2015, \mnras, 446, 663  
\bibitem[Halpern(1984)]{1984ApJ...281...90H} Halpern, J.~P.\ 1984,
  \apj, 281, 90 
\bibitem[Higginbottom et al.(2014)]{2014ApJ...789...19H} Higginbottom et al.\ 2014, \apj, 789, 19 
\bibitem[Kaastra et 
al.(2000)]{2000A&A...354L..83K} Kaastra et al.\ 2000, A\&A, 354, L83 
\bibitem[Kaastra et al.(2014)]{2014Sci...345...64K} Kaastra et al.\
  2014, Science, 345, 64
\bibitem[Kaastra et al.(2014)]{2014arXiv1412.1171K} Kaastra et al.\ 2014, arXiv:1412.1171
\bibitem[Kallman \& Bautista(2001)]{2001ApJS..133..221K} Kallman, T.,
  \& Bautista, M.\ 2001, \apjs, 133, 221  
\bibitem[Kaspi et al.(2002)]{2002ApJ...574..643K} Kaspi et al.\ 2002,
  \apj, 574, 643
\bibitem[Kazanas et al.(2012)]{2012AstRv...7c..92K} Kazanas et al.\
  2012, The Astronomical Review, 7, 92
\bibitem[King 
\& Pounds(2003)]{2003MNRAS.345..657K} King, A.~R., \& Pounds, K.~A.\
2003, \mnras, 345, 657  
\bibitem[King \& Pounds(2015)]{2015arXiv150305206K} King, A., \&
  Pounds, K.\ 2015, arXiv:1503.05206
\bibitem[Konigl \& Kartje(1994)]{1994ApJ...434..446K} Konigl, A., \&
  Kartje, J.~F.\ 1994, \apj, 434, 446 
\bibitem[Kormendy \& Ho(2013)]{2013ARA&A..51..511K} Kormendy, J., \&
  Ho, L.~C.\ 2013, \araa, 51, 511 
\bibitem[Lohfink et al.(2013)]{2013ApJ...772...83L} Lohfink et al.\
  2013, \apj, 772, 83  
\bibitem[Lobban et al.(2011)]{2011MNRAS.414.1965L} Lobban et al.\
  2011, \mnras, 414, 1965 
\bibitem[Magorrian et al.(1998)]{1998AJ....115.2285M} Magorrian et
  al.\ 1998, \aj, 115, 2285
\bibitem[Marscher et al.(2002)]{2002Natur.417..625M} Marscher et al.\
  2002, Nature, 417, 625   
\bibitem[McKernan et al.(2007)]{2007MNRAS.379.1359M} McKernan et
  al. 2007, \mnras, 379, 1359 
\bibitem[McKinney et al.(2014)]{2014MNRAS.441.3177M} McKinney et al. \
  2014, \mnras, 441, 3177
\bibitem[Nandra et al.(2013)]{2013arXiv1306.2307N} Nandra et al.\
  2013, arXiv:1306.2307  
\bibitem[Nardini et al.(2015)]{2015Sci...347..860N} Nardini et al.\ 2015, Science, 347, 860 
\bibitem[Ohsuga et al.(2009)]{2009PASJ...61L...7O} Ohsuga et al.\
  2009, \pasj, 61, L7 
\bibitem[Pounds et al.(2003)]{2003MNRAS.345..705P} Pounds et al.\ 2003, \mnras, 345, 705 
\bibitem[Pounds(2014)]{2014SSRv..183..339P} Pounds, K.\ 2014, SSRv,
  183, 339 
\bibitem[Proga(2000)]{2000ApJ...538..684P} Proga, D.\ 2000, \apj, 538,
  684 
\bibitem[Proga et al.(2000)]{2000ApJ...543..686P} Proga, D., Stone,
  J.~M., \& Kallman, T.~R.\ 2000, \apj, 543, 686 
\bibitem[Reeves et al.(2013)]{2013ApJ...776...99R} Reeves et al.\
  2013, \apj, 776, 99 
\bibitem[S{\c a}dowski et al.(2013)]{2013MNRAS.436.3856S} S{\c
    a}dowski et al. \ 2013, \mnras, 436, 3856 
\bibitem[S{\c a}dowski et al.(2014)]{2014MNRAS.439..503S} S{\c
    a}dowski et al. \ 2014, \mnras, 439, 503 
\bibitem[Silk \& Rees(1998)]{1998A&A...331L...1S} Silk, J., \& Rees,
  M.~J.\ 1998, A\&A, 331, L1 
\bibitem[Sim et al.(2008)]{2008MNRAS.388..611S} Sim et al.\ 2008, \mnras, 388, 611 
\bibitem[Sim et al.(2010)]{2010MNRAS.404.1369S} Sim et al.\ 2010, \mnras, 404,
  1369
\bibitem[Sim et al.(2012)]{2012MNRAS.426.2859S} Sim et al.\ 2012,
  \mnras, 426, 2859 
\bibitem[Takahashi et al.(2014)]{2014SPIE.9144E..25T} Takahashi et
  al.\ 2014, SPIE, 9144, 914425
\bibitem[Tchekhovskoy et al.(2011)]{2011MNRAS.418L..79T} Tchekhovskoy
  et al. \ 2011, \mnras, 418, L79
\bibitem[Tombesi et al.(2010)]{2010A&A...521A..57T} Tombesi et al.\ 2010a, A\&A, 521, A57
\bibitem[Tombesi et al.(2010)]{2010ApJ...719..700T} Tombesi et al.\
  2010b, \apj, 719, 700
\bibitem[Tombesi et al.(2011)]{2011ApJ...742...44T} Tombesi et al.\
  2011a, \apj, 742, 44
\bibitem[Tombesi et al.(2012)]{2012MNRAS.422L...1T} Tombesi et al.\
  2012a, \mnras, 422, L1 
\bibitem[Tombesi et al.(2012)]{2012MNRAS.424..754T} Tombesi et al.\
  2012b, \mnras, 424, 754 
\bibitem[Tombesi et al.(2013)]{2013MNRAS.430.1102T} Tombesi et al.\ 2013, \mnras, 430, 1102 
 \bibitem[Tombesi et al.(2014)]{2014MNRAS.443.2154T} Tombesi et al.\ 2014, \mnras, 443, 2154
\bibitem[Tombesi et al.(2015)]{2015Natur.519..436T} Tombesi et al.\ 2015, Nature, 519, 436 
\bibitem[Veilleux et al.(2013)]{2013ApJ...776...27V} Veilleux, S.,
  Mel{\'e}ndez, M., Sturm, E., et al.\ 2013, \apj, 776, 27
\bibitem[Wagner et al.(2013)]{2013ApJ...763L..18W} Wagner, A.~Y., et al.\ 2013, \apjl, 763, L18
\bibitem[Yuan \& Narayan(2014)]{2014ARA&A..52..529Y} Yuan, F., \&
  Narayan, R.\ 2014, \araa, 52, 529
\bibitem[Zubovas \& King(2012)]{2012ApJ...745L..34Z} Zubovas, K., \&
  King, A.\ 2012, \apjl, 745, L34 
\bibitem[Zubovas \& Nayakshin(2014)]{2014MNRAS.440.2625Z} Zubovas, K.,
  \& Nayakshin, S.\ 2014, \mnras, 440, 2625 
\end{thebibliography}
\end{document}